\documentclass[10pt,twocolumn,letterpaper]{article}

\usepackage{cvpr}
\usepackage{times}
\usepackage{epsfig}
\usepackage{graphicx}
\usepackage{amsmath}
\usepackage{amssymb}
\usepackage{array}
\usepackage{bm}
\usepackage{color}
\usepackage{algorithm}
\usepackage{algorithmic}
\usepackage[labelformat=simple]{subcaption}
\usepackage{multirow}
\usepackage{booktabs}
\usepackage{graphbox} 

\newcolumntype{C}[1]{>{\centering\arraybackslash}p{#1}}

\newenvironment{packed_enum}{
\begin{enumerate}
  \setlength{\itemsep}{1pt}
  \setlength{\parskip}{0pt}
  \setlength{\parsep}{0pt}
}{\end{enumerate}}

\usepackage[breaklinks=true,bookmarks=false]{hyperref}

\cvprfinalcopy %

\def\cvprPaperID{7604} %

\ifcvprfinal\fi

\makeatletter
\let\latexps@plain\ps@plain
\newcommand{\frontmatter}{\let\ps@plain\ps@empty\pagestyle{empty}}
\newcommand{\mainmatter}{%
  \let\ps@plain\latexps@plain\pagestyle{plain}%
  \clearpage
  \pagenumbering{arabic}}
\makeatother

\frontmatter

\begin{document}

\title{A Physics-based Noise Formation Model for Extreme Low-light Raw Denoising}

\author{Kaixuan Wei$^1$  \quad Ying Fu$^1$\footnote{Corresponding author: fuying@bit.edu.cn} \quad Jiaolong Yang$^2$  \quad Hua Huang$^1$
\\ $^1$Beijing Institute of Technology \quad $^2$Microsoft  Research\\
}

\maketitle

\begin{abstract}
  Lacking rich and realistic data, learned single image denoising algorithms
  generalize poorly to real raw images that do not resemble the data used for
  training. Although the problem can be alleviated by the
  heteroscedastic Gaussian model for noise synthesis, the noise sources caused by digital
  camera electronics are still largely overlooked, despite their significant effect on
  raw measurement, especially under extremely low-light condition. To address
  this issue, we present a highly accurate noise formation model based on the
  characteristics of CMOS photosensors, thereby enabling us to synthesize
  realistic samples that better match the physics of image formation process.
  Given the proposed noise model, we additionally propose a
  method to calibrate the noise parameters for available modern digital
  cameras, which is simple and reproducible for any new device. We
  systematically study the generalizability of a neural network trained with
  existing schemes, by introducing a new low-light denoising dataset that covers
  many modern digital cameras from diverse brands. Extensive empirical results
  collectively show that by utilizing our proposed noise formation model, a
  network can reach the capability as if it had been trained with rich real
  data, which demonstrates the effectiveness of our noise formation model.
\end{abstract}

\section{Introduction}
\renewcommand*{\thefootnote}{\fnsymbol{footnote}}
\stepcounter{footnote}\footnotetext{Corresponding author: fuying@bit.edu.cn}
\renewcommand*{\thefootnote}{\arabic{footnote}}
\setcounter{footnote}{0}
\begin{figure}[!htbp]
\centering
\begin{subfigure}[b]{.32\linewidth}
\centering
\includegraphics[width=1\linewidth,clip,keepaspectratio]{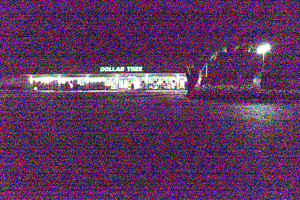}
\caption{Input}
\end{subfigure}
\begin{subfigure}[b]{.32\linewidth}
\centering
\includegraphics[width=1\linewidth,clip,keepaspectratio]{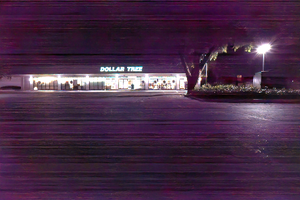}
\caption{G}
\end{subfigure}
\begin{subfigure}[b]{.32\linewidth}
\centering
\includegraphics[width=1\linewidth,clip,keepaspectratio]{./figures/images/sid/10111/sony-gp.png}
\caption{G+P}
\end{subfigure}
\begin{subfigure}[b]{.32\linewidth}
\centering
\includegraphics[width=1\linewidth,clip,keepaspectratio]{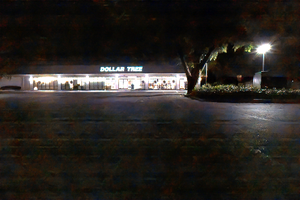}
\caption{Paired real data}
\end{subfigure}
\begin{subfigure}[b]{.32\linewidth}
\centering
\includegraphics[width=1\linewidth,clip,keepaspectratio]{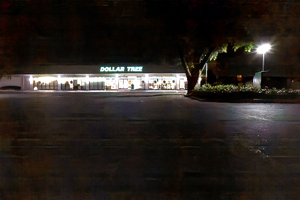}
\caption{Ours}
\end{subfigure}
\begin{subfigure}[b]{.32\linewidth}
\centering
\includegraphics[width=1\linewidth,clip,keepaspectratio]{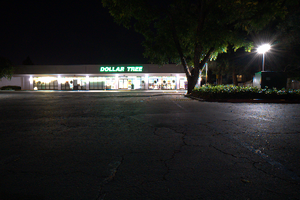}
\caption{Reference}
\end{subfigure} 
\caption{An image from the See-in-the-Dark (SID) Dataset \cite{Chen_2018_CVPR}, where we present (a) the short-exposure noisy input image; (f) the long-exposure reference image;  (b-e) the outputs of UNets \cite{ronneberger2015u} trained with (b) synthetic data generated by the  homoscedastic Gaussian noise model (G), (c) synthetic data generated by the signal-dependent heteroscedastic Gaussian noise model (G+P) \cite{Foi2008Practical},  (d) paired real data of \cite{Chen_2018_CVPR}, and (e) synthetic data generated by our proposed noise model respectively. \emph{All images were converted from raw Bayer space to sRGB for visualization; similarly hereinafter.}
}
\label{fig:example}
\end{figure}

Light is of paramount importance to photography. 
 Night and low light place very demanding
constraints on photography due to limited photon count and inescapable noise.
The natural reaction
is to gather more light by, \eg, enlarging aperture
setting, lengthening exposure time and opening flashlight. However, each method
is a tradeoff -- large aperture incurs small depth of field, and is unavailable
in smartphone cameras; long exposure can induce blur due to scene variations or
camera motions; flash can cause color aberrations and is useful only for
nearby objects. 

A practical rescue for low-light imaging is to use burst capturing
\cite{Mildenhall_2018_CVPR,Hasinoff2016Burst,Liu2014Fast,liba2019handheld}, in which a burst of
images are aligned and fused to increase the signal-to-noise ratio (SNR). However, 
burst photography can be fragile, suffering from ghosting effect
\cite{Hasinoff2016Burst,Shen_2019_ICCV} when capturing dynamic scenes in the presence of
vehicles, humans, \etc .
An emerging alternative approach is to employ a neural network to automatically
learn the mapping from a low-light noisy image to its long-exposure counterpart
\cite{Chen_2018_CVPR}. However, such a deep learning approach generally requires
a large amount of labelled training data that resembles low-light photographs in the real
world. 
Collecting rich
high-quality training samples from diverse modern camera devices is
tremendously labor-intensive and expensive.

In contrast, synthetic data is simple, abundant and inexpensive,
but its efficacy is highly contingent upon how accurate the adopted noise formation model is. 
The heteroscedastic Gaussian noise model \cite{Foi2008Practical}, instead of the
commonly-used homoscedastic one, approximates well the real noise occurred in
daylight or moderate low-light settings
\cite{Brooks2018Unprocessing,Shi2018Toward,Hasinoff2016Burst}. However, it cannot
delineate the full picture of sensor noise under severely low
illuminance. %
An illustrative example is shown in Fig.~\ref{fig:example}, where the
objectionable banding pattern artifacts, an unmodeled noise component that is
exacerbated in dim environments, become clearly noticeable by human eyes.

\begin{figure}[!t]
\centering
\includegraphics[width=1\linewidth,clip,keepaspectratio]{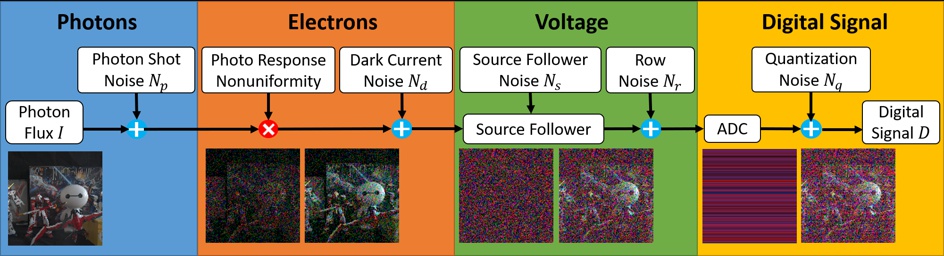}
\caption{Overview of electronic imaging pipeline and visualization of noise sources and the resulting image at each stage. }
\label{fig:photosensor}
\end{figure}

In this paper, to avoid the effect on noise model from the image processing pipeline (ISP)
\cite{Chen_2018_CVPR,Brooks2018Unprocessing,Mildenhall_2018_CVPR} converting raw data to
sRGB, we mainly focus on the noise formation model for raw images. We
propose a physics-based noise formation model for extreme low-light raw
denoising, which explicitly leverages the characteristics of CMOS photosensors
to better match the physics of noise formation. As shown in Fig.
\ref{fig:photosensor}, our proposed synthetic pipeline derives from the inherent
process of electronic imaging by considering how photons go through several
stages.
It models sensor noise in a fine-grained manner that includes many noise sources such as photon shot noise, pixel circuit noise, and quantization noise.
Besides, we
provide %
a method to calibrate the noise parameters from available digital
cameras.
In order to %
investigate the generality of our noise model, we additionally introduce
an extreme low-light denoising (ELD) dataset taken by various camera devices to
evaluate our model. Extensive experiments show that 
the network trained only with the synthetic data from our noise model can reach
the capability as if it had been trained with rich real data.

Our main contributions can be summarized as follows:
\vspace{-2mm}
\begin{itemize}
	\item We formulate a noise model to synthesize realistic noisy images that can match the quality of real data under extreme low-light conditions.\vspace{-3pt}
	\item We present a noise parameter calibration method that can adapt our model to a given camera.\vspace{-3pt}
	\item We collect a dataset with various camera devices to verify the effectiveness and generality of our model. 
\end{itemize}

\section{Related Work} \label{sec:related-work}

Noise removal from a single image is an extensively-studied yet still unresolved problem in
computer vision and image processing. Single image denoising methods generally
rely on the assumption that both signal and noise exhibit particular statistical
regularities such that they can be separated from a single observation. Crafting
an analytical regularizer associated with image priors (\eg smoothness,
sparsity, self-similarity, low rank), therefore, plays a critical role in
traditional design pipeline of denoising algorithms \cite{Rudin1992Nonlinear,osher2005iterative,elad2006image,dong2011sparsity,mairal2008sparse,dabov2007BM3D,Buades2005A,Gu2014Weighted}.
In the modern era, most single image denoising algorithms are entirely
data-driven, which consist of deep neural networks that implicitly learn the
statistical regularities to infer clean images from their noisy counterparts
\cite{Schmidt_2014_CVPR,Chen_2015_CVPR,mao2016image,zhang2017beyond,Gharbi:2016:DJD:2980179.2982399,tai2017memnet,Chen_2018_ECCV,Shi2018Toward}. Although simple and powerful, these
learning-based approaches are often trained on
synthetic image data due to practical constraints. The most widely-used additive, white, Gaussian noise model
deviates strongly from realistic evaluation scenarios, resulting in significant
performance declines on photographs with real noise \cite{Plotz_2017_CVPR,Abdelhamed_2018_CVPR}.
 
To step aside the domain gap between synthetic images and real photographs, some
works have resorted to collecting paired real data not just for evaluation but
for training \cite{Abdelhamed_2018_CVPR,Chen_2018_CVPR,Schwartz2018DeepISP,Chen_2019_ICCV,Jiang_2019_ICCV}.
Notwithstanding the promising results, collecting sufficient real data with ground-truth labels to prevent overfitting is exceedingly expensive and
time-consuming. %
Recent works exploit the use of paired (Noise2Noise
\cite{pmlr-v80-lehtinen18a}) or single (Noise2Void
\cite{Krull_2019_CVPR}) noisy images as training data instead
of paired noisy and noise-free images. However, they can not substantially ease the
burden of labor requirements for capturing a massive amount of real-world
training data.

Another line of research has focused on improving the realism of synthetic
training data to circumvent the difficulties in acquiring real data from
cameras. By considering both photon arrival statistics (``shot" noise) and sensor
readout effects (``read" noise), the works of \cite{Mildenhall_2018_CVPR,Brooks2018Unprocessing} employed a signal-dependent heteroscedastic Gaussian
model \cite{Foi2008Practical} to characterize the noise properties in raw sensor
data. Most recently, Wang \etal \cite{Wang_2019_ICCV} proposes a noise model, which considers the dynamic streak noise, color channel heterogeneous and clipping effect, to simulate the high-sensitivity noise on real low-light color images.  Concurrently, a flow-based generative model, namely Noiseflow \cite{Abdelhamed_2019_ICCV} is proposed to formulate the distribution of real noise using latent variables with tractable density\footnote{Note that Noiseflow requires paired real data to obtain noise data (by subtracting the ground truth images from the noisy ones) for training.}. 
However, these approaches %
oversimplify the modern
sensor imaging pipeline, especially the noise sources caused by camera
electronics, which have been extensively studied in the electronic
imaging community \cite{Mikhail2014Highlevel,Healey1994Radiometric,Gow2007A,Baer2006A,El2005CMOS,Farrell2008Sensor,Irie2008A,Irie2008A2,Boie1992An,Wach2004Noise,Costantini2004Virtual}. 
In this work, we propose a
physics-based noise formation model stemming from the essential process of
electronic imaging to synthesize the noisy dataset and show that sizeable improvements of denoising
performance on real data, particularly under extremely low illuminance.

\section{Physics-based Noise Formation Model} \label{sec:noise-model}
The creation of a digital sensor raw image $D$ can be generally formulated by a
linear model
  \begin{equation}
    \label{eq:formation}
    D = K I + N,
  \end{equation}
where $I$ is the number of photoelectrons that is proportional to the scene
irradiation, $K$ represents the overall system gain composed by analog and
digital gains, and $N$ denotes the summation of all noise sources physically caused
by light or camera.
We focus on  the single raw image denoising problem under extreme low-light
conditions. In this context, the characteristics of $N$ are formated in terms of
the sensor
physical process %
beyond the existing noise models.
Deriving an optimal regularizer to tackle such noise is infeasible, as
there is no analytical solver for such a noise distribution\footnote{Even if each noise component has an analytical formulation, their summation can generally be intractable.}. Therefore, we rely on a learning-based neural network
pipeline to implicitly learn the regularities from data. Creating training
samples for this task requires careful considerations of the characteristics of
raw sensor data. In the following, we first describe the detailed procedures of
the physical formation of a sensor raw image as well as the noise sources
introduced during the whole process. An overview of this process is shown in
Fig.~\ref{fig:photosensor}.

\subsection{Sensor Raw Image Formation} \label{sec: raw formation}
Our photosensor model is primarily based upon the CMOS sensor, which is the dominating imaging sensor nowadays \cite{grandviewresearch}. 
  We consider the electronic imaging pipeline of how incident light is converted from
photons to electrons, from electrons to voltage, and finally from voltage to
digital numbers, to model noise.

\vspace{3pt}
\noindent\textbf{From Photon to Electrons.~}
During exposure, incident lights in the form of photons hit the photosensor pixel area, which liberates photon-generated electrons
(photoelectrons) proportional to the light
intensity. 
Due to the quantum nature of light, there exists an inevitable
uncertainty in the number of electrons collected. %
Such uncertainty imposes a Poisson distribution over this number of electrons, which follows
\vspace{-1mm}
 \begin{align}
(I + N_p) \sim \mathcal{P} \left( I \right),
\label{eq: noise-poisson}
 \vspace{-1mm}
 \end{align}
where $N_p$ is termed as the \textit{photon shot noise} and $\mathcal{P}$ denotes the
Poisson distribution. This type of noise depends on the light intensity, \ie, on
the signal. Shot noise is a fundamental limitation and cannot be avoided even
for a perfect sensor. There are other noise sources introduced during the
photon-to-electron stage, such as photo response nonuniformity and dark current
noise, reported by many previous literatures \cite{Healey1994Radiometric,Gow2007A,Wach2004Noise,Baer2006A}. Over the last decade, technical advancements
in CMOS sensor design and fabrication, \eg, on-sensor dark current suppression,
have led to a new generation of digital single lens reflex (DSLR) cameras with lower dark current and better
photo response uniformity \cite{Fossum2014A,lin2016high}. Therefore, we assume a constant
photo response and absorb the effect of dark current noise $N_d$ into read noise $N_{read}$, which will be presented next.
 
\vspace{3pt}
\noindent\textbf{From Electrons to Voltage.~}
After electrons are collected at each site, they are typically integrated, amplified
and read out as measurable charge or voltage at the end of exposure time.
Noise present during the electrons-to-voltage stage depends on the circuit design and processing technology used, and thus is referred to as pixel circuit noise~\cite{Gow2007A}. It includes thermal noise, reset noise~\cite{Mikhail2014Highlevel}, source follower noise~\cite{Leyris2005Trap} and banding pattern noise~\cite{Gow2007A}. The physical origin of these noise components can be found in the electronic imaging literatures \cite{Mikhail2014Highlevel,Gow2007A,Wach2004Noise,Leyris2005Trap}. For instance, source follower noise is attributed to the action of traps in silicon lattice which randomly capture and emit carriers; banding pattern noise is associated with the CMOS circuit readout pattern and the amplifier.  %

By leveraging this knowledge,  we consider the thermal noise $N_{t}$, source follower noise $N_s$ and banding
pattern noise $N_{b}$ in our model.  The noise model of $N_{b}$ will be presented later. Here, 
we absorb multiple noise sources
into a
unified term, \ie read noise
\vspace{-1mm}
\begin{align}
N_{read} = N_d  + N_t +  N_s.
\vspace{-1mm}
\end{align}
\textit{Read noise} can be assumed to follow a Gaussian distribution, but the analysis of
noise data (in Section \ref{sec:noise-param}) tells a long-tailed nature of its
shape. This can be attributed by the flicker and random telegraph signal components of source
follower noise \cite{Gow2007A}, or the dark spikes raised by dark current
\cite{Mikhail2014Highlevel}. 
Therefore, we propose using a statistical distribution that can better characterize the long-tail shape.
Specifically, we model the read
noise by a Tukey lambda distribution ($TL$)~\cite{Joiner1971Some}, which is a
distributional family that can approximate a number of common distributions (\eg,
a heavy-tailed Cauchy distribution):
\vspace{-1mm}
\begin{align}
N_{read} \sim \mathop{TL} \left( \lambda; 0, \sigma_{TL} \right),
\label{eq: noise-tl}
\vspace{-1mm}
\end{align}
where $\lambda$ and $\sigma$ indicate the shape and scale parameters
respectively, while the location parameter is set to be zero given the
zero-mean noise assumption.

Banding pattern noise $N_b$ appears in images as horizontal or vertical lines. We only consider the row noise component (horizontal stripes) in our model, as the column noise component (vertical stripes) is generally negligible
when measuring the noise data (Section \ref{sec:noise-param}). We simulate the
\textit{row noise} $N_r$ by sampling a value from a zero-mean Gaussian distribution with a scale parameter $\sigma_r$, then adding it as an offset to the whole pixels within a single row. 

\vspace{3pt}
\noindent\textbf{From Voltage to Digital Numbers.~}
To generate an image that can be stored in a digital storage medium, the
analog voltage signal read out during last stage is quantized into discrete
codes using an ADC. This process introduces \textit{quantization noise} $N_q$ given by
\vspace{-1mm}
\begin{align}
N_q \sim U \left( -1/2 q , 1/2 q \right),
\vspace{-1mm}
\end{align}
where $U (\cdot, \cdot)$ denotes the uniform distribution over the range
$[-1/2 q , 1/2 q]$ and $q$ is the quantization step.

\textbf{To summarize}, our noise formation model consists of four major noise components:
\vspace{-1mm}
\begin{align}
N = K N_p + N_{read} + N_r + N_q,
\label{eq: noise-formation}
\vspace{-1mm}
\end{align}
where $K$, $N_p$, $N_{read}$, $N_r$ and $N_q$ denotes the overall system gain, photon shot noise, read noise, row noise and quantization noise, respectively. 

\subsection{Sensor Noise Evaluation} \label{sec:noise-param}

\begin{figure}[!t]
\centering
\begin{subfigure}[b]{.35\linewidth}
\centering
\includegraphics[width=1\linewidth,clip,keepaspectratio]{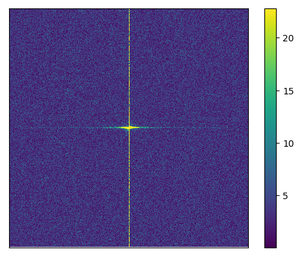}
\end{subfigure}
\begin{subfigure}[b]{.35\linewidth}
\centering
\includegraphics[width=1\linewidth,clip,keepaspectratio]{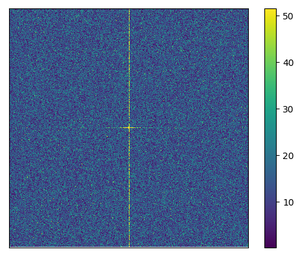}
\end{subfigure}
\vspace{-1mm}
\caption{Centralized Fourier spectrum of bias frames captured by SonyA7S2 (\textit{left}) and (\textit{right}) NikonD850 cameras}
\vspace{-3mm}
\label{fig:fft_bf}
\end{figure}

In this section, we present a noise parameter calibration method attached to our proposed noise formation model. %
According to Eq.~\eqref{eq: noise-poisson} \eqref{eq: noise-tl} \eqref{eq: noise-formation}, the necessary parameters to specify our noise model include
overall system gain $K$ for photon shot noise $N_p$; shape and scale parameters
($\lambda$ and $\sigma_{TL}$) for read noise $N_{read}$;  scale parameter $\sigma_r$ for row noise $N_r$. 
Given a new camera,  our noise calibration method consists of two main procedures, \ie (1) estimating noise parameters at various ISO settings\footnote{Noise parameters are generally stationary at a fixed ISO.}, and (2)  modeling joint distributions of noise parameters.

\vspace{3pt}
\noindent\textbf{Estimating noise parameters.~}
We record two sequences of raw images to estimate $K$ and other noise parameters: \emph{flat-field
frames} and \emph{bias frames}.

Flat-field frames are the images captured when sensor is uniformly illuminated.
They can be used to derive $K$ according to the Photon Transfer method.
\cite{Janesick1985CCD}%
Once we have $K$, we can firstly convert a raw digital signal $D$ into the number of photoelectrons $I$, then impose a Poisson distribution on it, and finally revert it to $D$ -- this simulates realistic photon shot noise.

\begin{figure}[!t]
\centering
\includegraphics[align=c,width=.32\linewidth,clip,keepaspectratio]{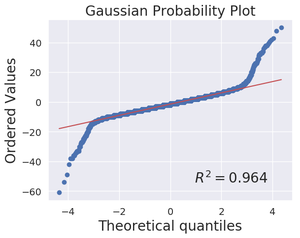}
\includegraphics[align=c,width=.32\linewidth,clip,keepaspectratio]{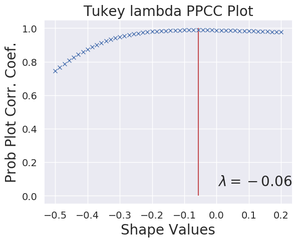}
\includegraphics[align=c,width=.32\linewidth,clip,keepaspectratio]{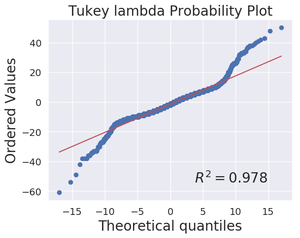} \\
\includegraphics[align=c,width=.32\linewidth,clip,keepaspectratio]{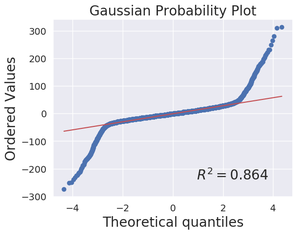}
\includegraphics[align=c,width=.32\linewidth,clip,keepaspectratio]{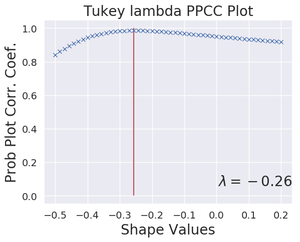}
\includegraphics[align=c,width=.32\linewidth,clip,keepaspectratio]{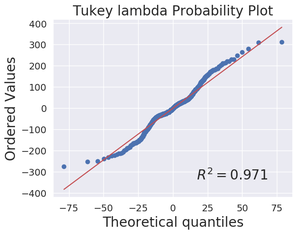} \\
\vspace{-1mm}
\caption{Distribution fitting of read noise for SonyA7S2 (\textit{top}) and  NikonD850 (\textit{bottom}) cameras. \textit{Left:} probability plot against the Gaussian distribution;
\textit{Middle:} Tukey lambda PPCC plot that determines the optimal $\lambda$ (shown in red line); \textit{Right:} probability plot against the Tukey Lambda distribution. A higher $R^2$ indicates a better fit. (\textbf{Best viewed with zoom})}
\vspace{0mm}
\label{fig:TL-PPCC}
\end{figure}

Bias frames are the images captured under a lightless environment with the
shortest exposure time. We took them at a dark room and the camera lens was capped on. 
Bias frames delineate the read noise picture independent of light, blended by the multiple noise sources aforementioned. 
The banding pattern noise can be tested via performing discrete Fourier transform on a bias frame.  In Fig.\ref{fig:fft_bf},  the highlighted vertical pattern in the centralized Fourier spectrum reveals the existence of row noise component. 
To analyze the distribution of row noise, we extract the
mean values of each row from raw data. These values, therefore, serve as good
estimates to the underlying row noise intensities, given the zero-mean nature
of other noise sources.  
The normality
of the row noise data is tested by a Shapiro-Wilk test \cite{Shapiro1975An}: the resulting $p$-value is higher than $0.05$, suggesting the null hypothesis that the data are normally distributed cannot be rejected. 
The related scale parameter $\sigma_r$ can be easily
estimated by maximizing the log-likelihood.

After subtracting the estimated row noise from a bias frame, 
statistical models can be used to fit the empirical distribution of the residual read noise. A preliminary diagnosis (Fig.~\ref{fig:TL-PPCC} Left) shows the main
body of the data may follow a Gaussian distribution, but it also
unveils the long-tail nature of the underlying distribution. 
In contrast to regarding extreme values as outliers, 
we observe an appropriate long-tail statistical distribution can
characterize the noise data better. 

We generate a probability plot correlation
coefficient (PPCC) plot~\cite{Filliben1975The} to identify a statistical model
from a Tukey lambda distributional family~\cite{Joiner1971Some} that best
describes the data. The Tukey lambda distribution is a family of distributions
that can approximate many distributions by varying its shape parameter
$\lambda$. It can 
approximate a Gaussian distribution if $\lambda = 0.14$,  or derive a
heavy-tailed distribution if $\lambda < 0.14$.  
 The PPCC plot (Fig.~\ref{fig:TL-PPCC} Middle) is used to find a good value of $\lambda$. The
probability plot~\cite{Wilk1968Probability} (Fig.~\ref{fig:TL-PPCC} Right) is then employed to estimate the scale parameter $\sigma_{TL}$. 
The goodness-of-fit can be evaluated by $R^2$ -- the
coefficient of determination \emph{w.r.t.} the resulting probability plot~\cite{MORGAN201115}. 
The $R^2$ of the fitted Tukey
Lambda distribution is much higher than the Gaussian distribution (\eg, $0.972$ \emph{vs.}
$0.886$), indicating a much better fit to the empirical data.

\begin{figure}[t]
	\centering
	\setlength\tabcolsep{1pt}
	\renewcommand\arraystretch{1}	
	\begin{tabular}{cccc}
		& \footnotesize Real Bias Frame & \footnotesize Gaussian Model & \footnotesize Ours \\
		\rotatebox[origin=c]{90}{\footnotesize SonyA7S2}	&
		\includegraphics[align=c,width=.32\linewidth,clip,keepaspectratio]{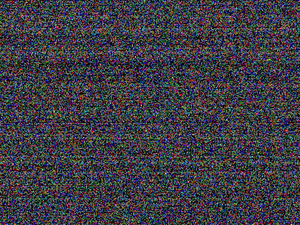} &
		\includegraphics[align=c,width=.32\linewidth,clip,keepaspectratio]{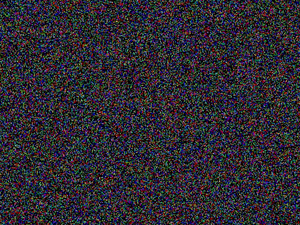} &
		\includegraphics[align=c,width=.32\linewidth,clip,keepaspectratio]{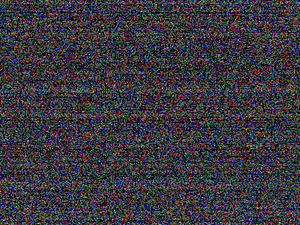} \\
		&  \footnotesize ($R^2$) & \footnotesize (0.961) & \footnotesize (0.978)\\
		\rotatebox[origin=c]{90}{\footnotesize NikonD850}	&
		\includegraphics[align=c,width=.32\linewidth,clip,keepaspectratio]{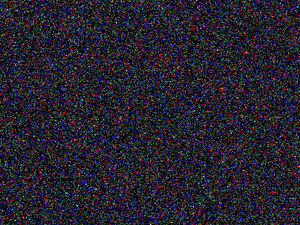} &
		\includegraphics[align=c,width=.32\linewidth,clip,keepaspectratio]{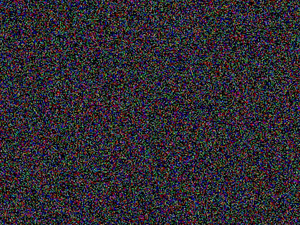} &
		\includegraphics[align=c,width=.32\linewidth,clip,keepaspectratio]{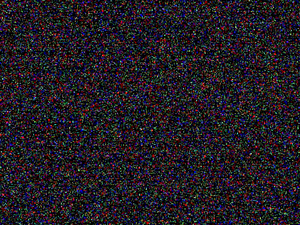} \\
		& \footnotesize ($R^2$) & \footnotesize (0.880) & \footnotesize (0.972) \\
	\end{tabular}
	\caption{Simulated and real bias frames of two cameras. A higher $R^2$ indicates a better fit quantitatively. \textbf{(Best viewed with zoom)}}\label{fig:noise_comparision}
\end{figure}

Although we use a unified noise model for different cameras,  
the noise parameters estimated from different cameras are highly diverse.  Figure~\ref{fig:TL-PPCC} shows the selected optimal shape parameter $\lambda$ differs camera by camera, implying distributions with varying degree of heavy tails across cameras.  The visual comparisons of real and simulated bias frames are shown in Fig.~\ref{fig:noise_comparision}. It shows that our model is capable of synthesizing realistic noise across various cameras, which outperforms the Gaussian noise model both in terms of the goodness-of-fit measure (\ie, $R^2$) and the visual similarity to real noise.

\begin{figure}[!t]
\centering
\begin{subfigure}[b]{.4\linewidth}
\centering
\includegraphics[width=1\linewidth,clip,keepaspectratio]{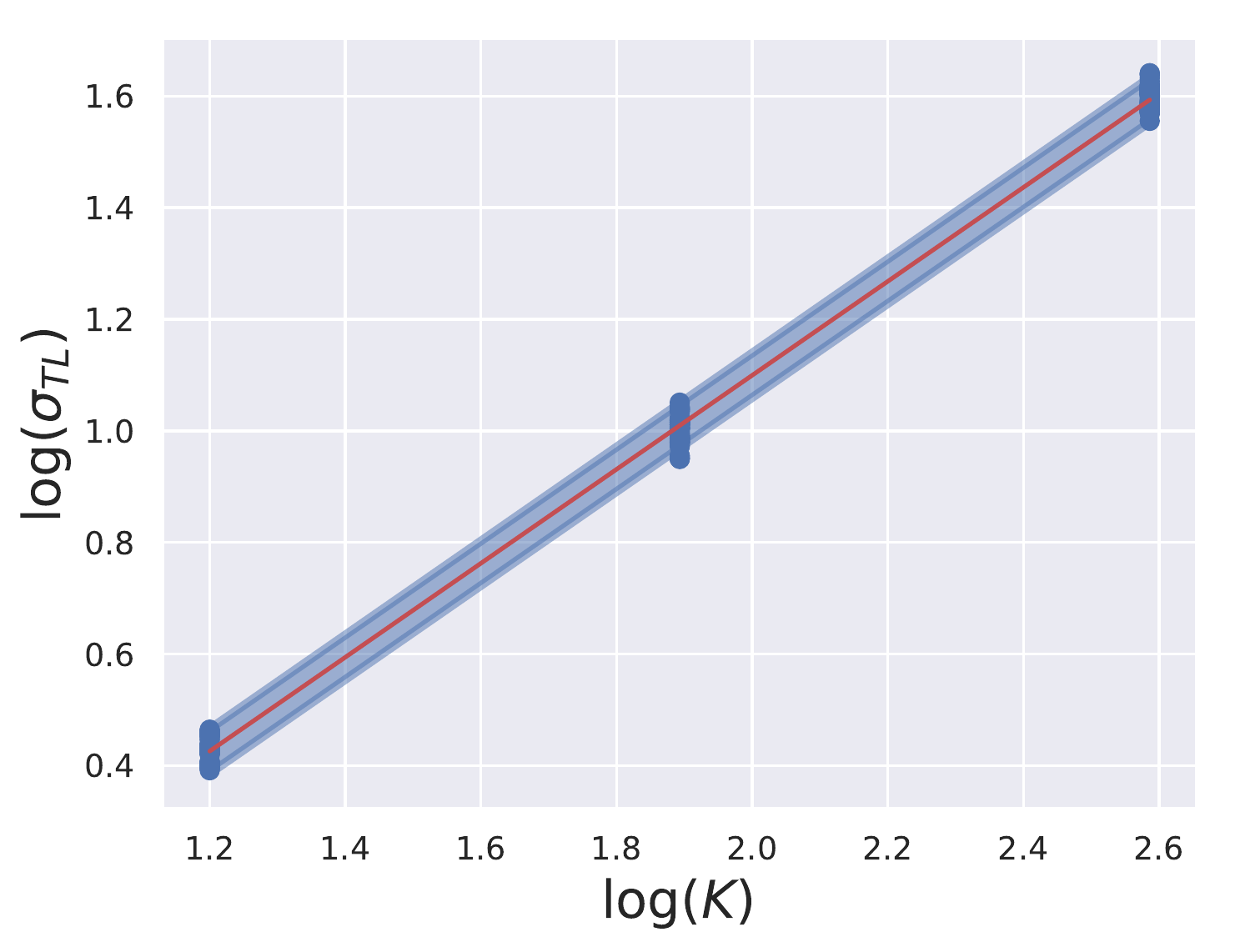}
\end{subfigure}
\begin{subfigure}[b]{.4\linewidth}
\centering
\includegraphics[width=1\linewidth,clip,keepaspectratio]{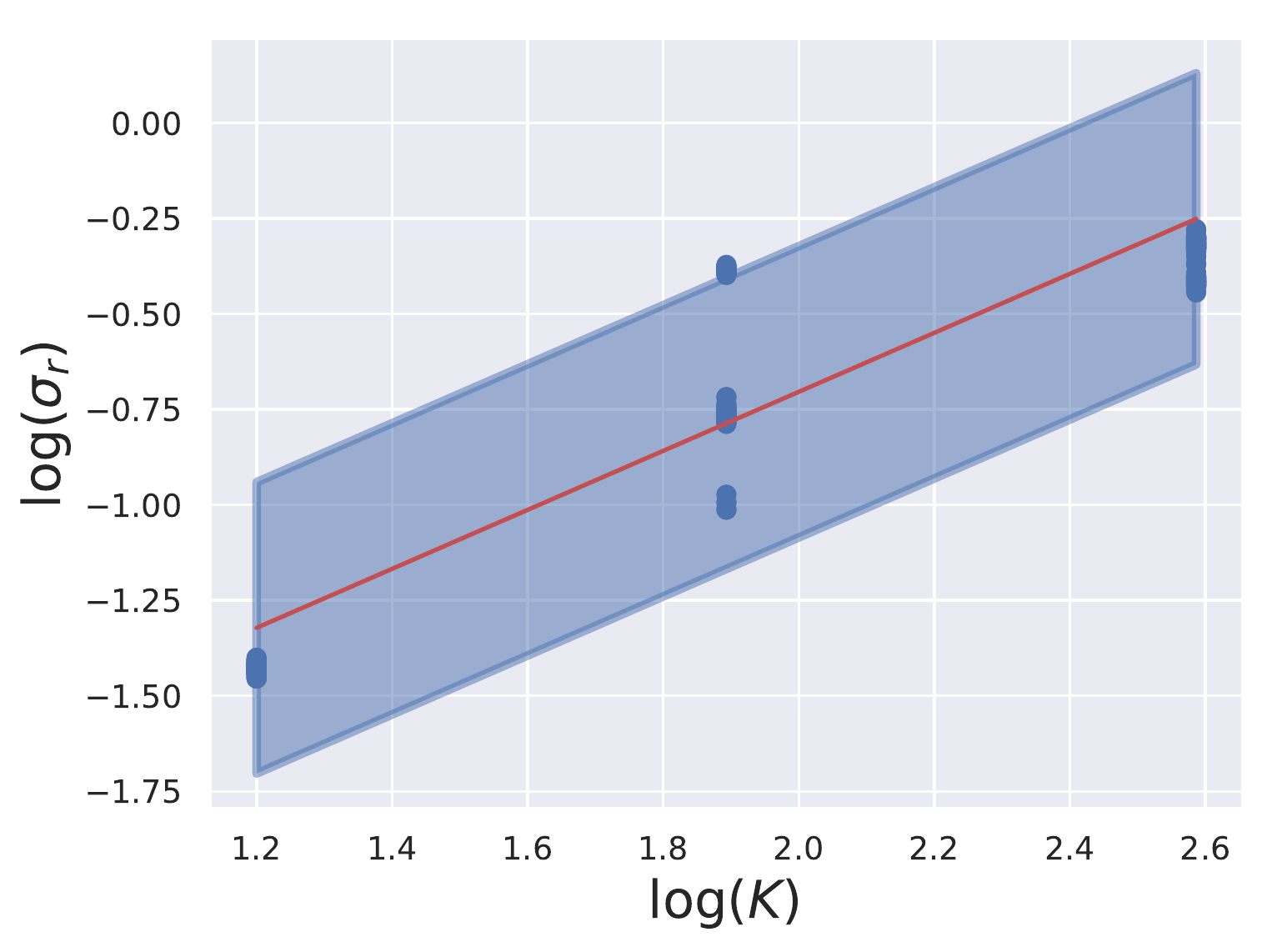}
\end{subfigure}
\caption{Linear least squares fitting from estimated noise parameter samples (blue dots) from a NikonD850 camera. Left and right figures show the joint distributions of $(K, \sigma_{TL})$ and $(K, \sigma_r)$ respectively, where we sample the noise parameters from the blue shadow regions. }
\label{fig:joint-dist}
\end{figure}

\vspace{3pt}
\noindent\textbf{Modeling joint parameter distributions.~}
To choose noise parameters for our noise formation model, we infer the
joint distributions of ($K$, $\sigma_{TL}$) and
($K$, $\sigma_{r}$), from the parameter samples estimated at various ISO settings.
As shown in Fig.~\ref{fig:joint-dist}, we use the linear least squares method to find the line of best fit for two sets of log-scaled measurements. Our noise parameter sampling procedure is
\vspace{-2mm}
\begin{align}
&\log \left( K \right)  \sim U \left( \log (\hat{K}_{min}), \log (\hat{K}_{max}) \right), \nonumber  \\   \label{eq: sampling}
&\log \left( \sigma_{TL} \right) | \log \left( K \right) \sim \mathcal{N} \left(a_{TL} \log (K)  + b_{TL} ,  \hat{\sigma}_{TL} \right), \\  
 &\log \left( \sigma_{r} \right) | \log \left( K \right) \sim \mathcal{N} \left(a_{r} \log (K)  + b_{r} ,  \hat{\sigma}_{r} \right),  \nonumber
 \vspace{-2mm}
\end{align}
where $U(\cdot,\cdot)$ denotes a uniform distribution and $\mathcal{N} (\mu, \sigma)$ denotes a Gaussian distribution with mean
$\mu$ and standard deviation $\sigma$. $\hat{K}_{min}$ and $\hat{K}_{max}$ are
the estimated overall system gains at the minimum and maximum ISO of a camera
respectively. $a$ and $b$ indicate the fitted line's slope and intercept respectively.
 $\hat{\sigma}$ is an unbiased
estimator of standard deviation of the linear regression under the Gaussian error
assumption. For shape parameter $\lambda$, we simply sample it from the
empirical distribution of the estimated parameter samples.

\vspace{3pt}
\noindent\textbf{Noisy image synthesis.~}
To synthesize noisy images, clean images are chosen and divided by low light factors sampled uniformly from $[100, 300]$ to simulate low photon count in the dark. Noise is then generated and added to
the scaled clean samples, according to Eq.~\eqref{eq: noise-formation}
\eqref{eq: sampling}. The created noisy images are finally normalized by multiplying the same low light factors to expose bright but excessively noisy contents.

\begin{figure}[!t]
\centering
\setlength\tabcolsep{0.8pt}
\begin{tabular}{cc}
\includegraphics[width=.45\linewidth,clip,keepaspectratio]{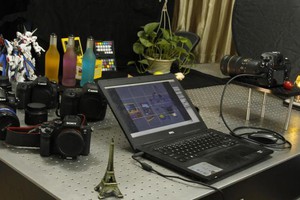} &
\includegraphics[width=.45\linewidth,clip,keepaspectratio]{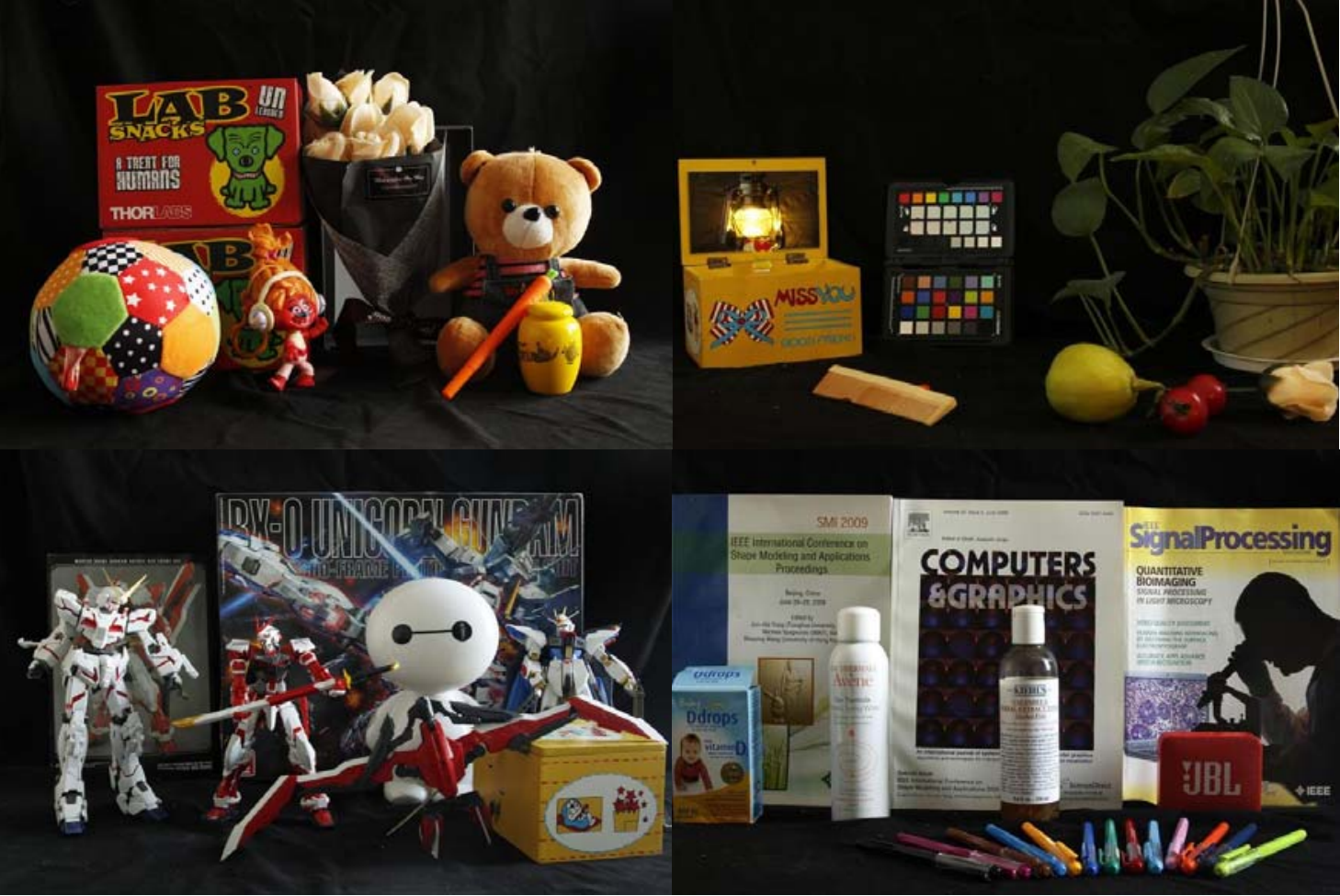} \\
\footnotesize (a) Image capture setup & \footnotesize (b) example images  \\
\end{tabular}
\caption{Capture setup and example images from our dataset.}
\label{fig:dataset}
\end{figure}

\begin{table}[!t]
\centering
\caption{Quantitative Results on Sony set of the SID dataset. The noise models are indicated as follows. $G$: the Gaussian model for read noise $N_{read}$; $G^*$: the tukey lambda model for $N_{read}$; $P$: the Gaussian approximation for photon shot noise $N_p$; $P^*$: the true Poisson model for $N_p$;  $R$: the Gaussian model for row noise $N_r$; $U$: the uniform distribution model for quantization noise $N_q$.
The best results are indicated by \textcolor{red}{red} color and the second best results are denoted by \textcolor{blue}{blue} color. }
\footnotesize
\begin{tabular}{lccc} 
		\toprule
		 & $\times 100$ & $\times 250$ & $\times 300$ \\
		Model & PSNR / SSIM & PSNR / SSIM & PSNR / SSIM \\ 
		\midrule
		BM3D  & 32.92 / 0.758 & 29.56 / 0.686 & 28.88 / 0.674 \\ \hline
		A-BM3D & 33.79 /  0.743 & 27.24 / 0.518 & 26.52 / 0.558 \\ \hline
		\midrule
		Paired real data  & 38.60 / \textcolor{blue}{0.912} & \textcolor{blue}{37.08} / \textcolor{red}{0.886} & \textcolor{blue}{36.29} / \textcolor{red}{0.874} \\ \hline
		Noise2Noise  & 37.42 / 0.853 & 33.48 / 0.725 & 32.37 / 0.686 \\ \hline
		\midrule
		$G$  & 36.10 / 0.800 & 31.87 / 0.640 & 30.99 / 0.624 \\ \hline
		$G$+$P$  & 37.08 / 0.839 & 32.85 / 0.697 & 31.87 / 0.665 \\ \hline
		$G$+$P^*$  & 38.31 / 0.884 & 34.39 / 0.765 & 33.37 / 0.730 \\ \hline
		$G^*$+$P^*$  & 39.10 / 0.911 & 36.46 / 0.869 & 35.69 / 0.855 \\ \hline
		$G^*$+$P^*$+$R$  & 39.23 / 0.912 & 36.89 / 0.877 & 36.01 / 0.864 \\ \hline 
		$G^*$+$P^*$+$R$+$U$  & \textcolor{red}{39.27} / \textcolor{red}{0.914} & \textcolor{red}{37.13} / \textcolor{blue}{0.883} & \textcolor{red}{36.30} / \textcolor{blue}{0.872} \\ \hline			
		
		\bottomrule
\end{tabular}
\label{tb:componet}
\end{table}

\section{Extreme Low-light Denoising (ELD) Dataset} \label{sec:eld-dataset}
To systematically study the generality of the proposed noise formation model, 
we collect an extreme low-light denoising (ELD) dataset that covers 10 indoor
scenes and 4 camera devices from multiple brands (SonyA7S2, NikonD850, CanonEOS70D, CanonEOS700D). %
We also record bias and flat field frames for each camera to calibrate our noise model. 
The data capture setup is shown in Fig.~\ref{fig:dataset}. 
For each scene and each camera, a reference image at the base ISO was firstly taken, followed by noisy images whose exposure time was deliberately decreased by low light factors $f$ to simulate extreme low light conditions.  Another reference image then was taken akin to the first one, to ensure no accidental error (\eg drastic illumination change or accidental camera/scene motion) occurred.
We  choose three ISO levels (800, 1600, 3200)\footnote{Most modern digital cameras are ISO-invariant when ISO is set higher than 3200 \cite{ISOless_Clark}.} and two low light factors (100, 200) for noisy images to capture our dataset, resulting in 240 (3$\times$2$\times$10$\times$4) raw image pairs in total. The hardest example in our dataset resembles the image captured at a ``pseudo" ISO up to 640000 (3200$\times$200).

\section{Experiments}

\subsection{Experimental setting} \label{sec:implementation-details}
\noindent\textbf{Implementation details.~}
A learning-based neural network pipeline is constructed to perform low-light raw denoising. We utilize the same U-Net architecture  \cite{ronneberger2015u} as \cite{Chen_2018_CVPR}.
Raw Bayer images from SID Sony training
dataset~\cite{Chen_2018_CVPR} are used to create training data. We pack the raw Bayer images into four channels (R-G-B-G) and crop non-overlapped $512\times 512$ regions augmented by random flipping/rotation.
Our approach only use the clean raw images, as the paired noisy images are generated by the proposed noise model on-the-fly.
Besides, we also train networks based upon other training schemes as references, including training with paired real data (short exposure and long exposure counterpart) and training with paired real noisy images (\ie, Noise2Noise~\cite{pmlr-v80-lehtinen18a}).

Our implementation\footnote{Code is released at https://github.com/Vandermode/NoiseModel} is based on PyTorch. %
We train the models with 200 epoch using $L_1$ loss and Adam optimizer~\cite{kingma2014adam} with batch size $1$. The learning rate is initially set to $10^{-4}$,   then halved at epoch 100, and finally reduced to $10^{-5}$ at epoch 180.

\vspace{3pt}
\noindent\textbf{Competing methods.~} To understand how accurate our proposed noise model is, we compare our method with:
\vspace{-1.8mm}
\begin{packed_enum}
\item The approaches that use real noisy data for training, \ie ``paired real data"  \cite{Chen_2018_CVPR}\footnote{\cite{Chen_2018_CVPR} used paired real data to perform raw-to-sRGB low-light image processing. Here we adapt its setting to raw-to-raw denoising.} and Noise2Noise \cite{pmlr-v80-lehtinen18a}; %
\item Previous noise models, \ie homoscedastic (G) and heteroscedastic Gaussian noise models (G+P) \cite{Foi2008Practical,foi2009clipped};
\item The representative non-deep methods, \ie BM3D \cite{dabov2007BM3D} and Anscombe-BM3D (A-BM3D) \cite{makitalo2011optimal}\footnote{The noise level parameters required are provided by the off-the-shelf image noise level estimators \cite{Foi2008Practical,Chen_2015_ICCV}.}.
\end{packed_enum}

\begin{figure}[t]
\vspace{-3.5mm}
	\centering
	\small
	\setlength\tabcolsep{1pt}
	\begin{tabular}{ccc}
			 &    &  \\
		 \includegraphics[width=0.3\linewidth]{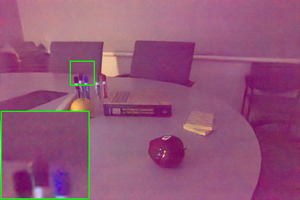}        
		& \includegraphics[width=0.3\linewidth]{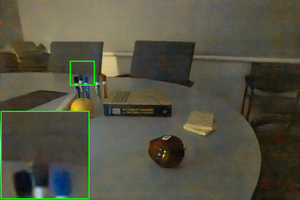} 
		 & \includegraphics[width=0.3\linewidth]{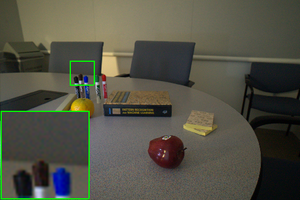}
 \\
		(a) Noise2Noise   &  (b)   Paired real data & (c)  Ground Truth  \vspace{2pt}\\
		 \includegraphics[width=0.3\linewidth]{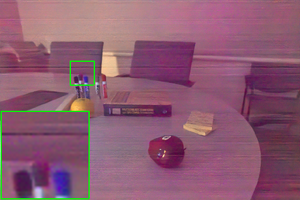} 
		 & \includegraphics[width=0.3\linewidth]{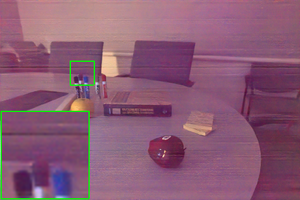} 
		 & \includegraphics[width=0.3\linewidth]{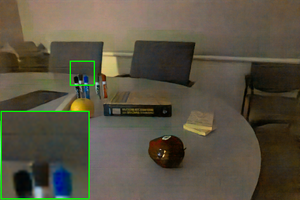} \\
		 (d) $G$  &  (e) $G$+$P$ & (f) $G^*$+$P^*$+$R$+$U$ \vspace{2pt}\\
	\end{tabular}
	\caption{Visual result comparison of different training schemes. Our final model ($G^*$+$P^*$+$R$+$U$) suppresses the ``purple" color shift, residual bandings and chroma artifacts compared to other baselines.}
	\label{fig:ablation-visual}
\end{figure}

\begin{table*}[!t]
	\centering
	\caption{Quantitative results (PSNR/SSIM) of different methods on our ELD dataset containing four representative cameras. }
	\footnotesize
	\setlength{\tabcolsep}{1mm}{
	\begin{tabular}{C{.11\linewidth}|C{.05\linewidth}|C{.05\linewidth}|C{.085\linewidth}|C{.09\linewidth}|C{.09\linewidth}|C{.11\linewidth}|C{.085\linewidth}|C{.085\linewidth}|C{.085\linewidth}}
		\hline
		\multirow{2}{*}{Camera} & \multirow{2}{*}{$f$} & \multirow{2}{*}{Index}  & \multicolumn{2}{c|}{Non-deep} & \multicolumn{2}{c|}{Training with real data} & \multicolumn{3}{c}{Training with synthetic data} \\ \cline{4-10}
		& & &  \!\!BM3D \cite{dabov2007BM3D}\!\! & \!\!A-BM3D~\cite{makitalo2011optimal}\!\! & \!\!Paired data~\cite{Chen_2018_CVPR}\!\!  & \!\!Noise2Noise~\cite{pmlr-v80-lehtinen18a}\!\! & \!\!G\!\! & \!\!G+P \cite{Foi2008Practical}\!\! & \!\!Ours\!\! \\\hline
\multirow{4}{*}{SonyA7S2}  & \multirow{2}{*}{$\times 100$} &PSNR&$37.69$&$37.74$&\textcolor{blue}{$44.50$}&$41.63$&$42.35$&$42.46$&\textcolor{red}{$45.36$}\\\cline{3-10}
& &SSIM&$0.803$&$0.776$&\textcolor{blue}{$0.971$}&$0.856$&$0.893$&$0.889$&\textcolor{red}{$0.972$}\\\cline{2-10}
& \multirow{2}{*}{$\times 200$} &PSNR&$34.06$&$35.26$&\textcolor{blue}{$42.45$}&$37.98$&$38.93$&$38.88$&\textcolor{red}{$43.27$}\\\cline{3-10}
& &SSIM&$0.696$&$0.721$&\textcolor{blue}{$0.945$}&$0.775$&$0.813$&$0.812$&\textcolor{red}{$0.949$}\\\hline
\multirow{4}{*}{NikonD850} & \multirow{2}{*}{$\times 100$} &PSNR&$33.97$&$36.60$&\textcolor{blue}{$41.28$}&$40.47$&$39.57$&$40.29$&\textcolor{red}{$41.79$}\\\cline{3-10}
& &SSIM&$0.725$&$0.779$&\textcolor{red}{$0.938$}&$0.848$&$0.823$&$0.845$&\textcolor{blue}{$0.912$}\\\cline{2-10}
& \multirow{2}{*}{$\times 200$} &PSNR&$31.36$&$32.59$&\textcolor{blue}{$39.44$}&$37.98$&$36.68$&$37.26$&\textcolor{red}{$39.69$}\\\cline{3-10}
& &SSIM&$0.618$&$0.723$&\textcolor{red}{$0.910$}&$0.820$&$0.757$&$0.786$&\textcolor{blue}{$0.875$}\\\hline
\multirow{4}{*}{CanonEOS70D} 
& \multirow{2}{*}{$\times 100$} &PSNR&$30.79$&$31.88$&$40.10$&$38.21$&$40.59$&\textcolor{red}{$40.94$}&\textcolor{blue}{$40.62$}\\\cline{3-10}
& &SSIM&$0.589$&$0.692$&$0.931$&$0.826$&$0.925$&\textcolor{blue}{$0.934$}&\textcolor{red}{$0.937$}\\\cline{2-10}
& \multirow{2}{*}{$\times 200$} &PSNR&$28.06$&$28.66$&$37.32$&$34.33$&$37.49$&\textcolor{blue}{$37.64$}&\textcolor{red}{$38.17$}\\\cline{3-10}
& &SSIM&$0.540$&$0.597$&$0.867$&$0.704$&$0.871$&\textcolor{blue}{$0.873$}&\textcolor{red}{$0.890$}\\\hline
\multirow{4}{*}{CanonEOS700D} 
&\multirow{2}{*}{$\times 100$} &PSNR&$29.70$&$30.13$&$39.05$&$38.29$ &$39.77$&\textcolor{red}{$40.08$}&\textcolor{blue}{$39.84$}\\\cline{3-10}
&&SSIM&$0.556$&$0.640$&\textcolor{blue}{$0.906$}&$0.859$&$0.884$&$0.897$&\textcolor{red}{$0.921$}\\\cline{2-10}
&\multirow{2}{*}{$\times 200$} &PSNR&$27.52$&$27.68$&$36.50$&$34.94$&\textcolor{blue}{$37.67$}&\textcolor{red}{$37.86$}&$37.59$\\\cline{3-10}
&&SSIM&$0.537$&$0.579$&$0.850$&$0.766$&$0.870$&\textcolor{blue}{$0.879$}&\textcolor{red}{$0.879$}\\\hline
	\end{tabular}}
	\label{tb:ELD}
\end{table*}

\begin{figure*}[!t]
	\centering
	\setlength\tabcolsep{1pt}
	\begin{tabular}{cccccccc}		
	Input & BM3D & A-BM3D & G & G+P & Noise2Noise & Paired real data & Ours   \\
	\includegraphics[width=0.12\linewidth]{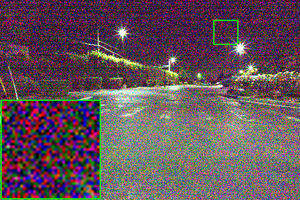}
	&		\includegraphics[width=0.12\linewidth]{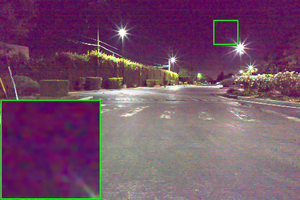}
	&		\includegraphics[width=0.12\linewidth]{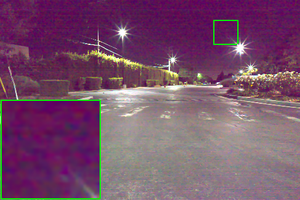}
	&		\includegraphics[width=0.12\linewidth]{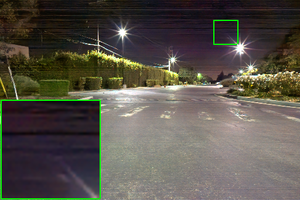}
	&	\includegraphics[width=0.12\linewidth]{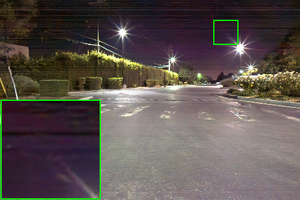}
	&		\includegraphics[width=0.12\linewidth]{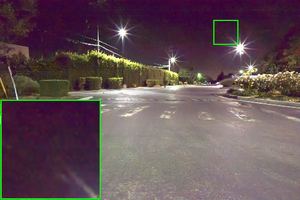}
	&	\includegraphics[width=0.12\linewidth]{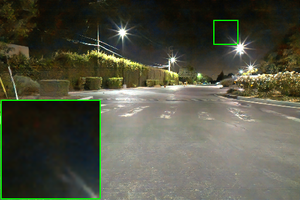}
	&	\includegraphics[width=0.12\linewidth]{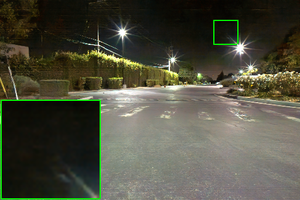} \\
	\includegraphics[width=0.12\linewidth]{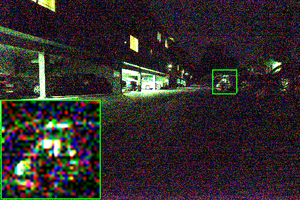}
	&		\includegraphics[width=0.12\linewidth]{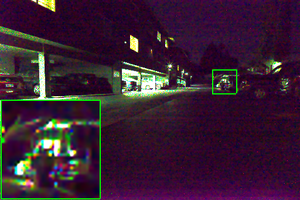}
	&		\includegraphics[width=0.12\linewidth]{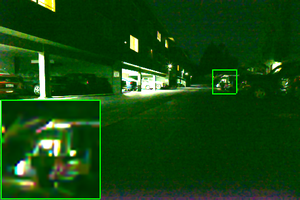}
	&		\includegraphics[width=0.12\linewidth]{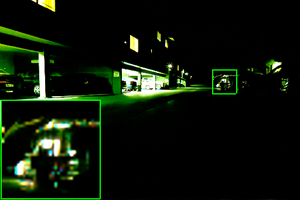}
	&	\includegraphics[width=0.12\linewidth]{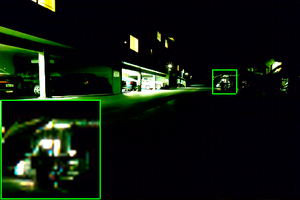}
	&		\includegraphics[width=0.12\linewidth]{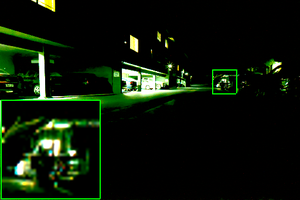}
	&	\includegraphics[width=0.12\linewidth]{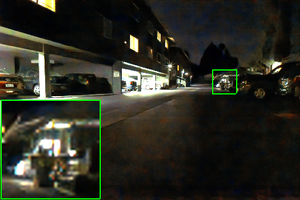}
	&	\includegraphics[width=0.12\linewidth]{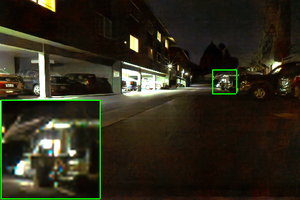} \\
	\includegraphics[width=0.12\linewidth]{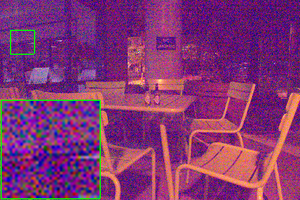}
	&		\includegraphics[width=0.12\linewidth]{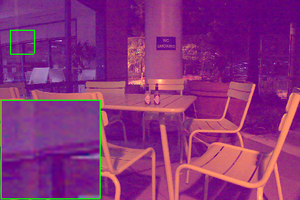}
	&		\includegraphics[width=0.12\linewidth]{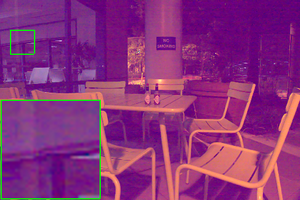}
	&		\includegraphics[width=0.12\linewidth]{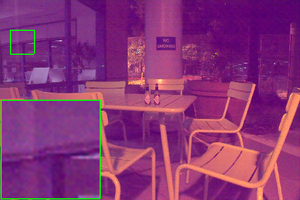}
	&	\includegraphics[width=0.12\linewidth]{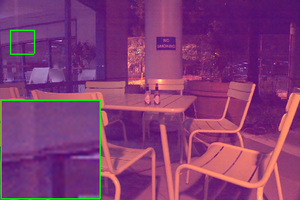}
	&		\includegraphics[width=0.12\linewidth]{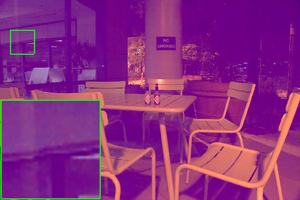}
	&	\includegraphics[width=0.12\linewidth]{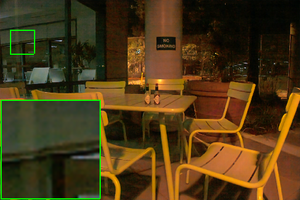}
	&	\includegraphics[width=0.12\linewidth]{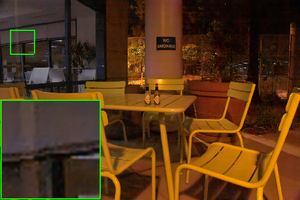} \\
	\end{tabular} 
	\caption{Raw image denoising results on both indoor and outdoor scenes from SID Sony dataset. (\textbf{Best viewed with zoom}) 
	}
	\label{fig:sony-vis}
\end{figure*}

\begin{figure*}[!t]
	\centering
	\setlength\tabcolsep{1pt}
	\begin{tabular}{cccccccc}
		Input  & BM3D & A-BM3D & G & G+P & Noise2Noise & Paired real data & Ours  \\
        	 \includegraphics[width=0.12\linewidth]{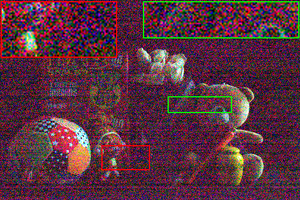}
        &	 \includegraphics[width=0.12\linewidth]{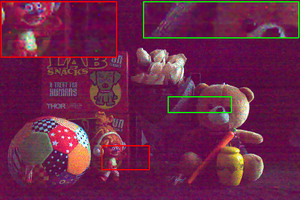}        
		&  \includegraphics[width=0.12\linewidth]{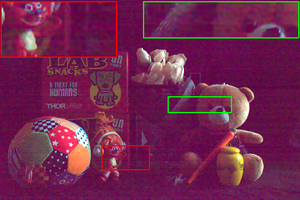}
		&  \includegraphics[width=0.12\linewidth]{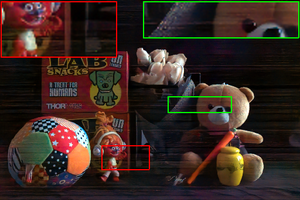}
		&  \includegraphics[width=0.12\linewidth]{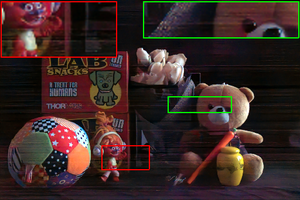}
		&	 \includegraphics[width=0.12\linewidth]{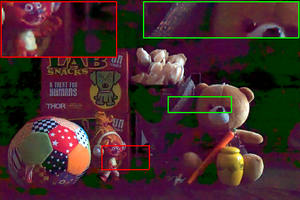}
		&  \includegraphics[width=0.12\linewidth]{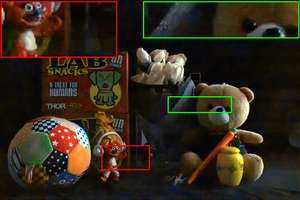}
		&  \includegraphics[width=0.12\linewidth]{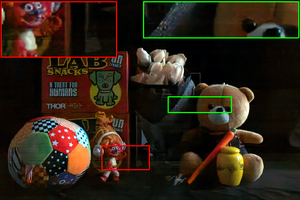} \\	
        	 \includegraphics[width=0.12\linewidth]{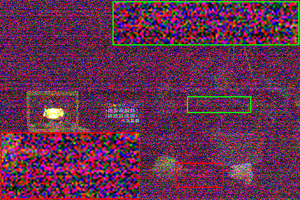}
		&	 \includegraphics[width=0.12\linewidth]{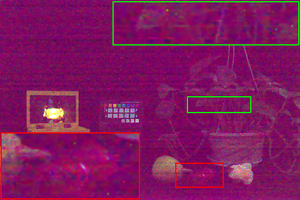}        	 
		&  \includegraphics[width=0.12\linewidth]{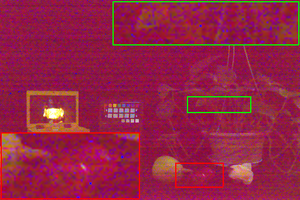}
		&  \includegraphics[width=0.12\linewidth]{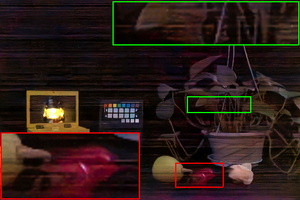}
		&  \includegraphics[width=0.12\linewidth]{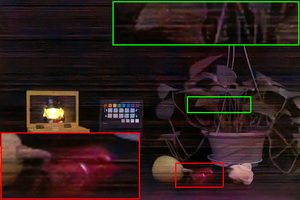}
		&	 \includegraphics[width=0.12\linewidth]{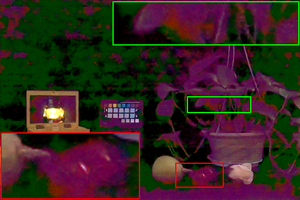}
		&  \includegraphics[width=0.12\linewidth]{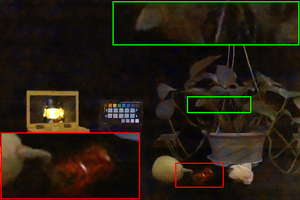}
		&  \includegraphics[width=0.12\linewidth]{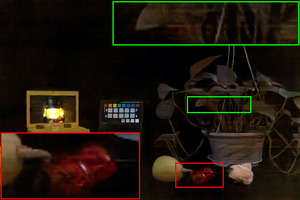} \\
        	 \includegraphics[width=0.12\linewidth]{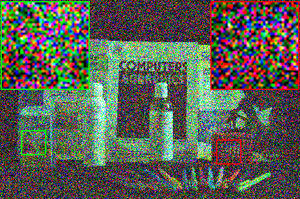}
	    &	 \includegraphics[width=0.12\linewidth]{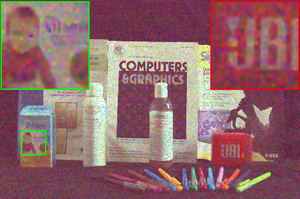}        	 
		&  \includegraphics[width=0.12\linewidth]{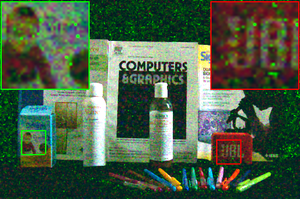}
		&  \includegraphics[width=0.12\linewidth]{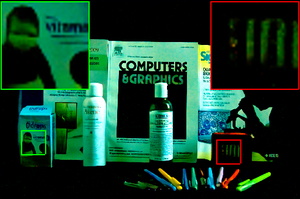}
		&  \includegraphics[width=0.12\linewidth]{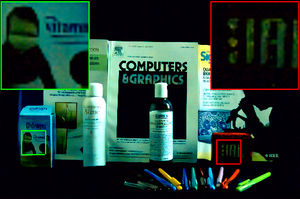}
		&	 \includegraphics[width=0.12\linewidth]{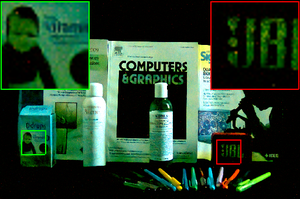}
		&  \includegraphics[width=0.12\linewidth]{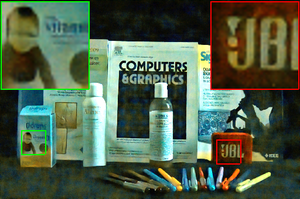}
		&  \includegraphics[width=0.12\linewidth]{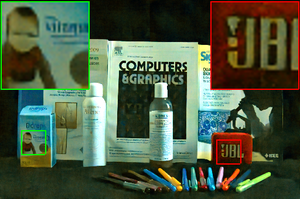}\\
        	 \includegraphics[width=0.12\linewidth]{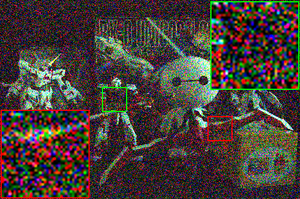}
		&	 \includegraphics[width=0.12\linewidth]{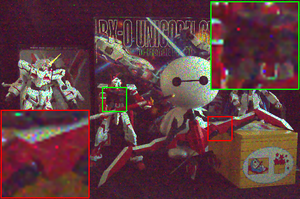}        	 
		&  \includegraphics[width=0.12\linewidth]{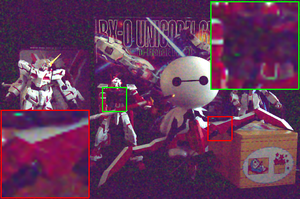}
		&  \includegraphics[width=0.12\linewidth]{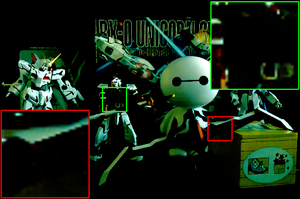}
		&  \includegraphics[width=0.12\linewidth]{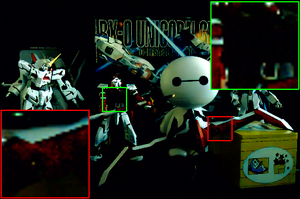}
		&	 \includegraphics[width=0.12\linewidth]{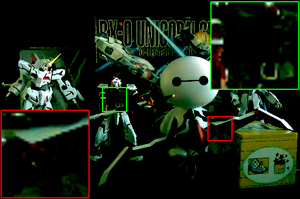}
		&  \includegraphics[width=0.12\linewidth]{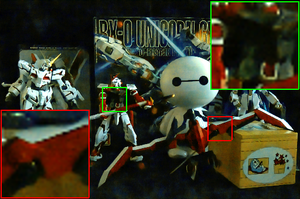}
		&  \includegraphics[width=0.12\linewidth]{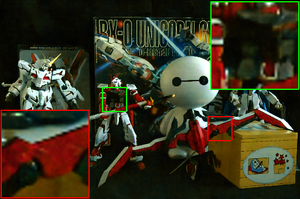}\\	
	\end{tabular} 
	\caption{Raw image denoising results on our ELD dataset. (\textbf{Best viewed with zoom)}}  %
	\label{fig:method-comparision}
\end{figure*}

\subsection{Results on SID Sony dataset}
Single image raw denoising experiment is firstly conducted on images from SID Sony validation and test sets.
For quantitative evaluation, we focus on indoor scenes illuminated by natural lights, to avoid flickering effect of alternating current lights \cite{Abdelhamed_2018_CVPR}
\footnote{Alternating current light is not noise, but a type of
illumination that breaks the irradiance constancy between short/long exposure pairs, making the quantitative evaluation inaccurate.}. To account for the imprecisions of shutter speed and analog gain \cite{Abdelhamed_2018_CVPR}, a single scalar
is calculated and multiplied into the reconstructed image to
minimize the mean square error evaluated by the ground
truth.

\vspace{3pt}
\noindent\textbf{Ablation study on noise models.}
To verify the efficacy of the proposed noise model, we compare the performance of networks trained with different noise models developed in Section \ref{sec: raw formation}.
All noise parameters are calibrated using the ELD
dataset, and sampled with a process following (or similar to) Eq.~\eqref{eq: sampling}.
The results of the other methods described in Section~\ref{sec:implementation-details} are also presented as references.

As shown in Table~\ref{tb:componet}, the domain gap is significant between the
homoscedastic/heteroscedastic Gaussian models and the de facto noise model
(characterized by the model trained with paired real data). \textit{This can be
attributed to (1) the Gaussian approximation of Possion distribution is not
justified under extreme low illuminance; (2) horizontal bandings are not
considered in the noise model; (3) long-tail nature of read noise is overlooked. }
By taking all these factors into account, our final model, \ie $G^*$+$P^*$+$R$+$U$ gives
rise to a striking result: the result is comparable to or sometimes even better
than the model trained with paired real data. 
Besides, training only with real low-light noisy data is not effective enough,
due to the clipping effects (that violates the zero-mean noise assumption) and
the large variance of corruptions (that leads to a large variance of the
Noise2Noise solution)~\cite{pmlr-v80-lehtinen18a}. A visual comparison of our final model and other methods is presented in Fig.~\ref{fig:ablation-visual}, which shows the effectiveness of our noise formation model.

Though we only quantitatively evaluate the results on indoor scenes of the SID Sony set,  %
our method can be applied to outdoor scenes as well. 
The visual comparisons of both indoor and outdoor scenes from SID Sony set are presented in Fig.~\ref{fig:sony-vis}. 
It can be seen that the random noise can be suppressed by the model learned with heteroscedastic Gaussian noise (G+P)~\cite{Foi2008Practical},  but the resulting colors are distorted, the banding artifacts become conspicuous, and the image details are barely discernible.  By contrast, our model produces visually appealing results as if it had been trained with paired real data.

\begin{figure}[!t]
	\centering
	\begin{subfigure}[b]{.32\linewidth}
		\centering
		\includegraphics[width=1\linewidth,clip,keepaspectratio]{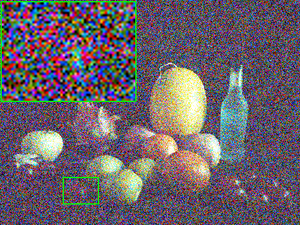}
		\caption{Input}
	\end{subfigure}
	\begin{subfigure}[b]{.32\linewidth}
		\centering
		\includegraphics[width=1\linewidth,clip,keepaspectratio]{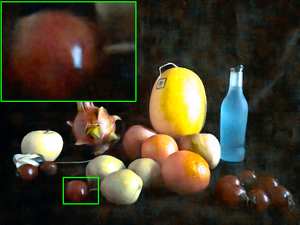}
		\caption{Paired real data}
	\end{subfigure}
	\begin{subfigure}[b]{.32\linewidth}
		\centering
		\includegraphics[width=1\linewidth,clip,keepaspectratio]{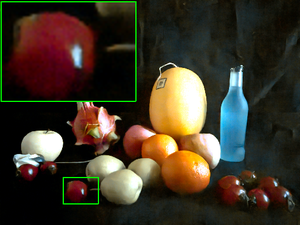}
		\caption{Ours}
	\end{subfigure}
	\caption{Denoising results of a low-light image captured by a Huawei Honor 10 camera.}
	\label{fig:smartphone}
\end{figure}

\begin{figure}[!t]
	\centering
	\begin{subfigure}[b]{.45\linewidth}
		\centering
		\includegraphics[width=1\linewidth,clip,keepaspectratio]{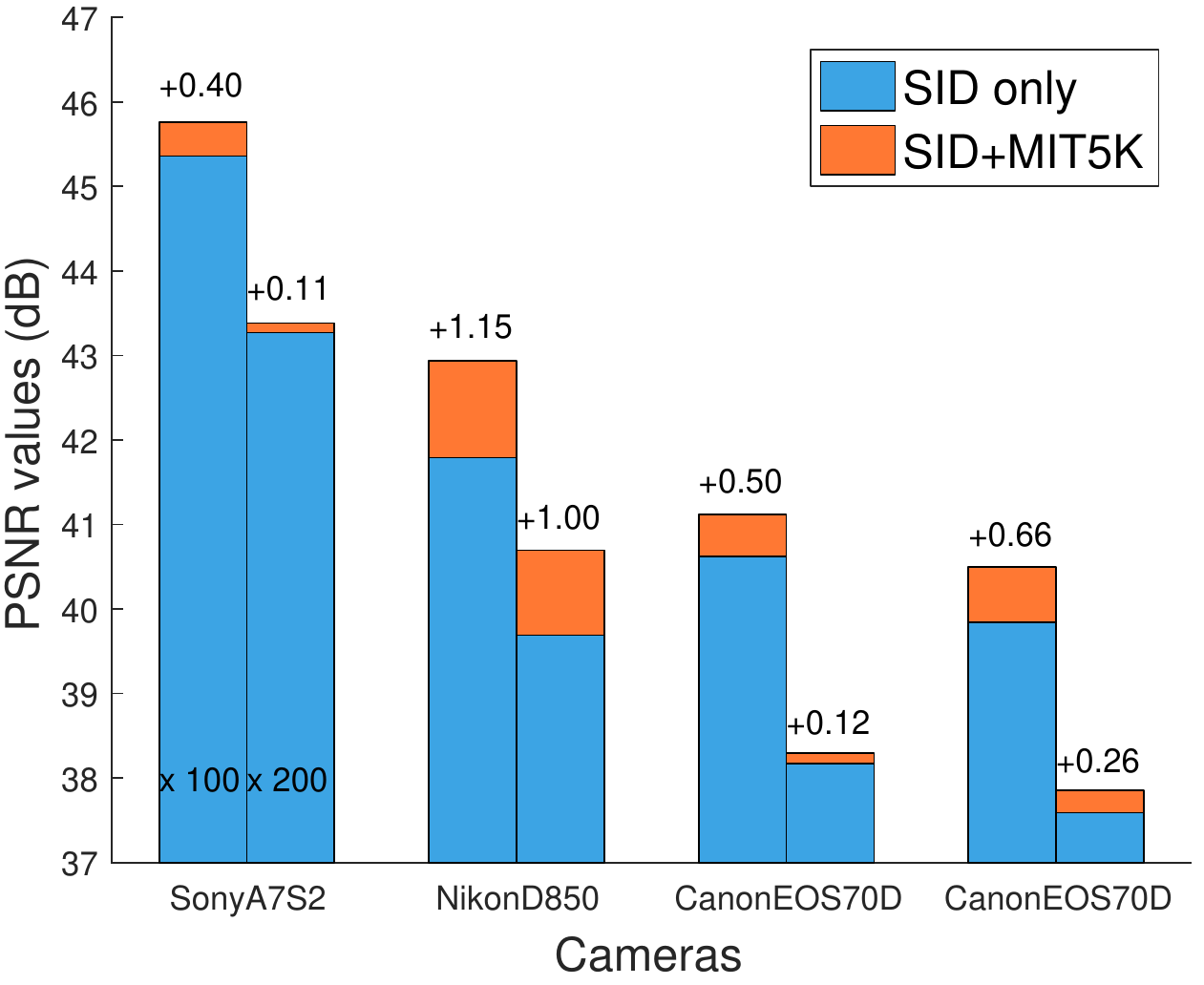}
		\caption{}
		\label{fig:mit5k}
	\end{subfigure}
	\begin{subfigure}[b]{.45\linewidth}
		\centering
		\includegraphics[width=1\linewidth,clip,keepaspectratio]{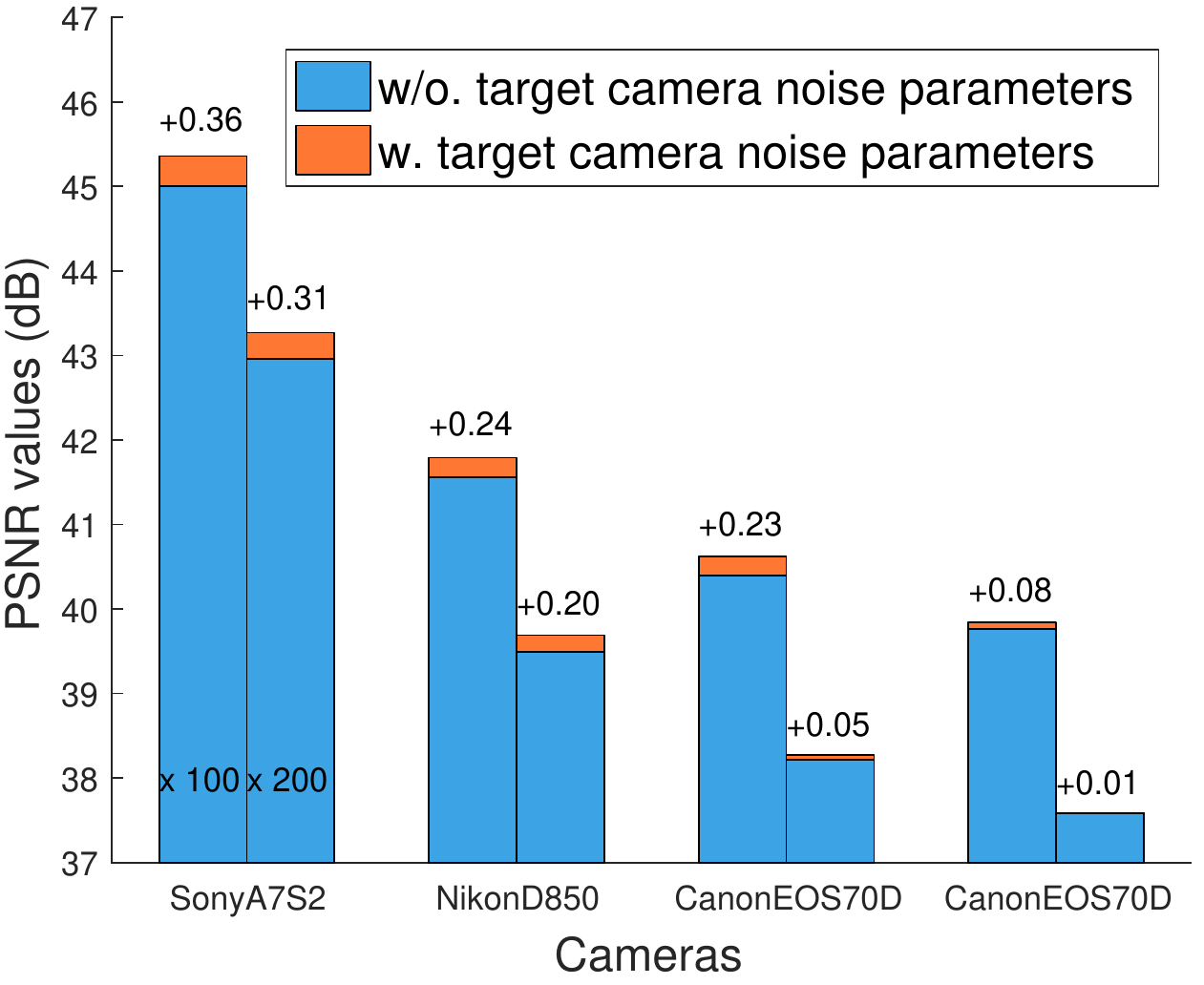}
		\caption{}
		\label{fig: parameter-sensitivity}
	\end{subfigure}
	\vspace{-2mm}
	\caption{(a) Performance boost when training with more synthesized data. (b) Noise parameter sensitivity test. }
	\vspace{-3mm}
\end{figure}

\subsection{Results on our ELD dataset}
\noindent\textbf{Method comparisons.~}
To see whether our noise model can be applicable to other camera devices as
well, we assess model performance on our ELD dataset.
Table~\ref{tb:ELD} and Fig.~\ref{fig:method-comparision} summarize the
results of all competing methods.  
It can be seen that the non-deep denoising methods, \ie BM3D and A-BM3D, fail to address the banding residuals, the color bias and the extreme values presented in the noisy input,  whereas our model recovers vivid image details which can be hardly perceived on the noisy image by human observers.
\textit{Moreover, our model trained with synthetic data even often outperforms the model trained with paired real data.} 
We note the finding here conforms with the evaluation of sensor noise presented in Section~\ref{sec:noise-param}, especially in Fig.~\ref{fig:TL-PPCC} and \ref{fig:noise_comparision}, where we show the underlying noise distribution varies camera by camera. 
Consequently, training with paired real data from SID Sony camera inevitably overfits to the noise pattern merely existed on the Sony camera, 
leading to suboptimal results on other types of cameras.  In contrast, our model relies on a very flexible noise model and a noise calibration process,  making it adapts to noise characteristics of other (calibrated) camera models as well.  Additional evidence can be found in Fig.~\ref{fig:smartphone}, where we apply these two models to an image captured by a smartphone camera. 
Our reconstructed image is clearer and cleaner than what is restored by the model trained with paired real data.  

\begin{figure}[!t]
\centering
\begin{subfigure}[b]{.32\linewidth}
\centering
\includegraphics[width=1\linewidth,clip,keepaspectratio]{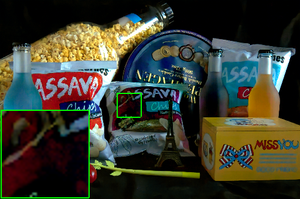}
\caption{SID only}
\end{subfigure}
\begin{subfigure}[b]{.32\linewidth}
\centering
\includegraphics[width=1\linewidth,clip,keepaspectratio]{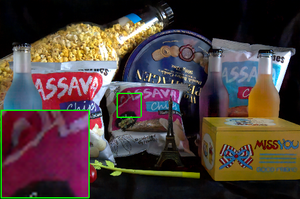}
\caption{SID + MIT5K}
\end{subfigure}
\begin{subfigure}[b]{.32\linewidth}
\centering
\includegraphics[width=1\linewidth,clip,keepaspectratio]{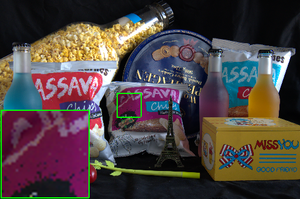}
\caption{Ground Truth}
\end{subfigure}
\caption{Denoising results of a low-light image captured by a NikonD850 camera. }
\label{fig:mit5k-vis}
\end{figure}

\vspace{3pt}
\noindent\textbf{Training with more synthesized data.~}
A useful merit of our approach against the conventional training with paired
real data, is that our model can be easily incorporated with more real clean
samples to train. Fig.~\ref{fig:mit5k} shows the relative improvements of our
model when training with the dataset synthesized by additional clean raw images from MIT5K dataset \cite{fivek}. We find the major improvements, as shown in Fig.~\ref{fig:mit5k-vis}, are owing to the more accurate color and brightness restoration. By training with more raw image samples from diverse cameras, the network learns to infer picture appearances more naturally and precisely. 

\vspace{3pt}
\noindent\textbf{Sensitivity to noise calibration.~}
Another benefit of our approach is we only need clean samples and a noise calibration process to adapt to a new camera, in contrast to capturing real noisy images accompanied with densely-labeled ground truth. 
Besides, the noise calibration process can be simplified once we already have a collection of parameter samples from various cameras. Fig.~\ref{fig: parameter-sensitivity} shows models can reach comparable performance on target cameras without noise calibration, by simply sampling parameters from other three calibrated cameras instead. 

\section{Conclusion} \label{sec:conclusion}
We have presented a physics-based noise formation model together with a noise parameter calibration method to help resolve the difficulty of extreme low-light denoising.  
We revisit the electronic imaging pipeline and investigate the influential noise sources overlooked by existing noise models.  
This enables us to synthesize realistic noisy raw data that better match the underlying physical process of noise formation. 
We systematically study the efficacy of our noise formation model by introducing a new dataset that covers four representative camera devices. 
By training only with our synthetic data, we demonstrate a convolutional neural network can compete with or sometimes even outperform the network trained with paired real data.

\vspace{6pt}
\noindent\textbf{Acknowledgments~}
We thank Tianli Tao for the great help in collecting the ELD dataset. This work was partially supported by the National Natural Science Foundation of China under Grants No. 61425013 and No. 61672096.

\end{document}